\documentclass[
superscriptaddress,twocolumn,
amsmath,amssymb,
aps,
]{revtex4-2}

\usepackage{bm}
\usepackage{amsmath}
\usepackage{amssymb}
\usepackage[T1]{fontenc}
\usepackage{array}
\usepackage{float}
\usepackage{graphicx}
\usepackage{color}
\usepackage{mathtools}
\usepackage{xcolor}

\makeatletter
\renewcommand{\fnum@figure}{\textbf{Fig. \thefigure}}
\makeatother
\newcommand{\argmax}[1]{\underset{#1}{\operatorname{arg}\,\operatorname{max}}\;}
\newcommand{\ket}[1]{\left|#1\right\rangle}
\newcommand{\bra}[1]{\left\langle #1\right|}

\newcommand{\const}[1]{\mathrm{#1}}

\begin{document}


\title{Determination of the asymptotic limits of adaptive photon counting measurements for coherent-state optical phase estimation}

\author{M. A. Rodríguez-García}
\affiliation{Instituto de Investigaciones en Matemáticas Aplicadas y en Sistemas, Universidad Nacional Autónoma de México, Ciudad Universitaria, Ciudad de. México 04510, Mexico}
		
\author{M. T. DiMario}
\affiliation{Center for Quantum Information and Control, Department of Physics and Astronomy, University of New Mexico, Albuquerque, New Mexico 87131}
\affiliation{Joint Quantum Institute, National Institute of Standards and Technology and the University of Maryland, College Park, Maryland 20742}

\author{P. Barberis-Blostein}
\affiliation{Instituto de Investigaciones en Matemáticas Aplicadas y en Sistemas, Universidad Nacional Autónoma de México, Ciudad Universitaria, Ciudad de. México 04510, Mexico}

\author{F. E. Becerra}
\email{fbecerra@unm.edu}
\affiliation{Center for Quantum Information and Control, Department of Physics and Astronomy, University of New Mexico, Albuquerque, New Mexico 87131}


\begin{abstract}

  Physical realizations of the canonical phase measurement for the optical phase
  are unknown. Single-shot phase estimation, which aims to determine the phase
  of an optical field in a single shot, is critical in quantum information
  processing and metrology. Here we present a family of strategies for
  single-shot phase estimation of coherent states based on adaptive
  non-Gaussian, photon counting, measurements with coherent displacements that
  maximize information gain as the measurement progresses, which have higher
  sensitivities over the best known adaptive Gaussian strategies. To gain
  understanding about their fundamental characteristics and demonstrate their
  superior performance, we develop a comprehensive statistical analysis based on
  Bayesian optimal design of experiments, which provides a natural description
  of these non-Gaussian strategies. This mathematical framework, together with
  numerical analysis and Monte Carlo methods, allows us to determine the
  asymptotic limits in sensitivity of strategies based on photon counting
  designed to maximize information gain, which up to now had been a challenging
  problem. Moreover, we show that these non-Gaussian phase estimation strategies
  have the same functional form as the canonical phase measurement in the
  asymptotic limit differing only by a scaling factor, thus providing the
  highest sensitivity among physically-realizable measurements for single-shot
  phase estimation of coherent states known to date. This work shines light into
  the potential of optimized non-Gaussian measurements based on photon counting
  for optical quantum metrology and phase estimation.
\end{abstract}

\maketitle

\section{\label{sec:Introduction}Introduction}
Optical phase estimation is ubiquitous in many fundamental and practical
problems ranging from quantum state preparation \cite{Giovann2004, DiMario2020,
  rafaldemkowiczdobrzanski12}, sensing \cite{PhysRevLett.116.061102},
communications \cite{Helstrom1969, becerra15,
  dolinar,PhysRevResearch.3.013200,PhysRevResearch.2.023384,dimario2019optimized,PhysRevLett.121.023603,DiMario:18,ferdinand2017multi,becerra2013experimental},
and information processing \cite{slussarenko2019photonic}. In a photonic
metrology problem
\cite{plenio05,polino2020photonic,giovannetti2011advances,Flamini_2018}, an
optical probe state interacts with a physical system to interrogate its
properties. This interaction maps parameters of the system to the state of the
optical probe, where an optimal readout can be performed
\cite{polino2020photonic, giovannetti2011advances, slussarenko2019photonic,
  Flamini_2018}. When the physical property of the system is mapped onto the
phase of the optical probe, the optimal quantum measurement is the canonical
phase measurement \cite{Wiseman1995}, which consists of projections onto phase
eigenstates \cite{wiseman98}. However, while theoretically the canonical phase
measurement exists, the physical realization of projections onto phase
eigenstates are not physically known \cite{Martin2019}. Thus in practical
estimation problems in quantum metrology one seeks to determine the limits in
precision of physically realizable measurements, and the degree with which they
approach to the fundamental quantum limit in sensitivity \cite{braunstein94,
  dobrzanski15, genoni11, lee19, bradshaw18, polino2020photonic}.

Physically realizable measurements of the optical phase have been widely
investigated \cite{Wiseman1995,wiseman97,wiseman98} for diverse metrological
problems with quantum and classical fields
\cite{anisimov10,Huang2017,anderson17,izumi16,slussarenko17} including sensing
small deviations from a known phase \cite{anisimov10,Huang2017,anderson17,
  anderson17b,izumi16,slussarenko17} and estimation with repeated sampling
\cite{daryanoosh18, higgins07} and measurements \cite{Huang2017, hentschel10,
  hou19, larson17, lumino18, zheng19}. Beyond these specific estimation
problems, measurements of the phase of a single optical mode in a single-shot
are central for quantum state preparation and detection
\cite{DEUTSCH2010681,bouchoule02}, waveform estimation and sensing
\cite{iwasawa13,tsang13,aasi13,nair12}, and quantum information processing
\cite{vanloock08,munro05,nemoto04}. The standard measurement for optical phase
estimation is the heterodyne measurement \cite{wiseman98}, which samples both
quadratures of the electromagnetic field simultaneously from which the phase can
be estimated. However, the achievable sensitivity of the heterodyne measurement
\cite{wiseman98} is far below the ultimate measurement sensitivity allowed by
physics, given by the canonical phase measurement \cite{Holevo2011,wiseman98}.
Adaptive measurement techniques based on homodyne detection, a Gaussian
measurement, can be used to align the phase quadrature of the optical field with
the measurement setting where the homodyne measurement provides maximum
sensitivity \cite{izumi16}. Adaptive homodyne has been theoretically shown to
surpass the heterodyne limit and get closer to the canonical phase measurement
for optical phase estimation of coherent states \cite{Wiseman1995,wiseman98},
providing the most sensitive Gaussian measurement of the optical phase so far
\cite{wiseman98}.

In a complementary measurement paradigm, quantum measurements of coherent states
based on photon counting, displacement operations, and feedback
\cite{PhysRevResearch.3.013200,DiMario2020,PhysRevResearch.2.023384,dimario2019optimized,PhysRevLett.121.023603,DiMario:18,ferdinand2017multi,becerra15,becerra2015photon,becerra2013experimental}
have enabled state discrimination below the Gaussian sensitivity limits and
approaching the Helstrom bound \cite{Helstrom1969}. Recently, some of the
authors proposed and demonstrated a non-Gaussian measurement strategy for
single-shot phase estimation of coherent states, able to surpass the heterodyne
limit and approach the sensitivity limit of a canonical phase measurement in the
presence of loss and noise of real systems \cite{DiMario2020}. These measurement
strategies are based on realizing coherent displacements of the input field and
monitoring the output field with photon number resolving detection. The
information from the detection outcomes is then used to implement real time
feedback of displacement operations optimized to maximize the measurement
sensitivity of the phase of the input state. This estimation strategy is the
most sensitive single-shot non-Gaussian measurement of a completely unknown
phase encoded in optical coherent states so far \cite{DiMario2020}. In this
strategy the displacement operation optimization is realized by maximizing
either the information gain in subsequent adaptive steps of the measurement or
the sharpness of the posterior phase distribution. While these cost functions
are functionally different, both perform similarly and get close to the ultimate
sensitivity allowed by physics, the Crámer-Rao lower bound (CRLB), in the limit
of many photons and many adaptive measurements. While the work in Ref.
\cite{DiMario2020} demonstrated the potential of non-Gaussian measurements for
single-shot phase estimation, the superiority over adaptive homodyne detection
was not proven. A deeper understanding of the properties of convergence and
ultimate limits of the estimators produced by non-Gaussian measurements is still
missing. This is an open problem due to the complexity associated with the
analysis of these adaptive non-Gaussian strategies.

In this article we investigate a family of adaptive non-Gaussian strategies
based on photon counting for single-shot optical phase estimation, and assess
their performance compared to Gaussian measurements and the canonical phase
measurement. To analyze these non-Gaussian strategies, we use the mathematical
framework of Bayesian optimal design of experiments, which provides a natural
description of non-Gaussian adaptive strategies, allowing us to investigate
their fundamental characteristics and determine the limits in sensitivity, which
up to now has been a challenging problem. Our work provides a comprehensive
statistical analysis of adaptive non-Gaussian measurements for parameter
estimation, and the requirements to approach optimal bounds in the asymptotic
limit. We show that strategies based on photon counting and feedback for
single-shot phase estimation of coherent states provide superior sensitivity
over the best known adaptive Gaussian strategies, having the same functional
form as the canonical phase measurement in the asymptotic limit, differing only
by a scaling factor. This work provides a deep insight into the potential of
optimized non-Gaussian measurements for quantum communication, metrology,
sensing, and information processing.


\section{\label{sec:Results} Results}

\subsection{Holevo Variance of Non-Gaussian Estimation Strategies}

The non-Gaussian phase estimation strategies investigated here are based on
photon counting, displacement operations, and feedback, and are optimized by
maximizing a specific cost function. These strategies maximize either the
estimation precision (by minimizing the Holevo variance \cite{Holevo2011}), or
the information gain about the unknown parameter based on entropy measures
including mutual information, the Kullback-Lieber divergence, and the
conditional entropy \cite{Cover2006,Chaloner1995}. We note that these cost
functions produce non-identifiable likelihood functions that do not allow to
correctly estimate a cyclic parameter, such as the phase
\cite{RodriguezGarcia2021efficientqubitphase}. To address this problem, these
non-Gaussian strategies use the Fisher information to optimize the displacement
operations, which are the dynamical control variable in the strategy, to
guarantee that these cost functions provide identifiable likelihood functions,
and to enable optical phase estimation with near optimal performance.

In the problem of single-shot phase estimation with coherent states, an
electromagnetic field in a coherent state $\rho_{0} = \ket{\alpha}\bra{\alpha}$
interacts with a physical system and experiences a unitary transformation
$e^{\const{i} \phi \hat{n}}$, where $\hat{n}$ is the number operator. The phase
$\phi$ induced in the coherent state carries information about the system, which
can be extracted by a measurement of the output state $\rho(\phi) = e^{\const{i}
  \phi \hat{n}} \rho_{0} e^{-\const{i} \phi \hat{n}}= \ket{e^{\const{i} \phi }
  \alpha}\bra{e^{-\const{i} \phi }\alpha} \,$.

Measurements onto $\rho(\phi)$, together with an estimator
$\widehat{\phi}$ on the measurement outcomes, provide an estimate of
$\phi$, and a measurement strategy aims to obtain the best estimation
of the physical parameter. The efficiency of such a
  strategy is characterized by the estimation variance as a function
  of the number of photons in the coherent state. The most
  efficient strategy provides a variance at the highest convergence
  rate towards zero as the number of photons increase.

The standard measurement paradigm for phase estimation of Gaussian states is the
heterodyne measurement (a Gaussian measurement), with an estimator variance of
$\text{Var}\left[ \widehat{\phi}_{\text{Het}} \right]=1/2 \lvert \alpha \rvert^2$. Within
the paradigm of Gaussian measurements, adaptive homodyne strategies optimized to
minimize the Holevo variance have achieved the best performance among Gaussian
measurements for single-shot phase estimation of coherent states
\cite{Wiseman1995}. The best adaptive Gaussian measurement reported to date,
termed the Adaptive Mark-II (MKII), achieves a Holevo variance in the limit of
large number of photons of:
\begin{equation}
  \label{eq:MKII_var}
 \text{Var}\left[ \widehat{\phi}_{\text{MKII}} \right] \approx \frac{1}{4 \lvert \alpha  \rvert^2} + \frac{1}{ 8 \lvert \alpha  \rvert^3}\, .
\end{equation}
While this optimized Gaussian strategy surpasses the heterodyne limit, it has an
error of order $1/|\alpha|^3$ above the Cramér-Rao Lower Bound ($1/4 \lvert
\alpha \rvert^2$), and does not reach the performance of the canonical phase
measurement \cite{wiseman98}:
\begin{equation}
  \label{eq:CPM_var}
 \text{Var}\left[ \widehat{\phi}_{\text{CPM}} \right] \approx \frac{1}{4 \lvert \alpha  \rvert^2} + \frac{5}{ 32 \lvert \alpha  \rvert^4}\,.
\end{equation}

\begin{figure*}[htbp!]
\centering
  \includegraphics[width=\textwidth]{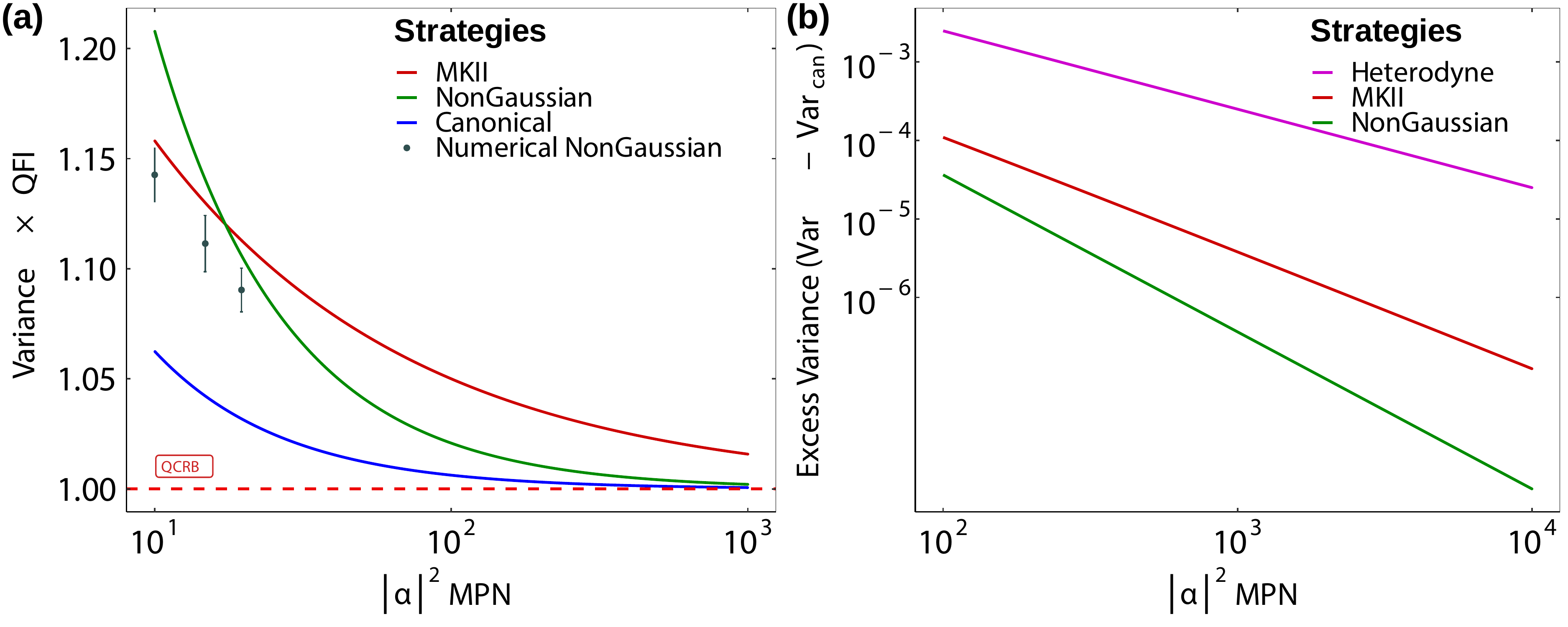}\quad
  \caption{ \textbf{Asymptotic variances for different phase measurements.}
    \textbf{a}: Holevo variance in the limit of large mean photon (MPN) $\lvert
    {\alpha} \rvert^2$ for the canonical phase measurement in Eq.
    (\ref{eq:CPM_var}), the most sensitive Gaussian measurement known to date
    (MKII) \cite{wiseman98} in Eq. (\ref{eq:MKII_var}), and the non-Gaussian
    strategies in Eq. (\ref{eq:model}). The points show numerically values for
    the non-Gaussian strategies in a region where the analytical expression is
    not valid. Error bars represent a 1-$\sigma$ standard deviation. \textbf{b}:
    Excess phase variance ($\text{Var}[\cdot]$ - $\text{Var}[
    \widehat{\phi}_{\text{CPM}} ]$) for Heterodyne, MKII, and non-Gaussian
    strategies.}
  \label{fig:Excess}
\end{figure*}

In this work, we numerically show that non-Gaussian strategies for single-shot
phase estimation based on photon counting, optimized displacement operations,
and real-time feedback achieve an estimator variance smaller than Gaussian
strategies with an asymptotic scaling in the limit of high mean photon numbers
of:
\begin{equation}
  \label{eq:model}
  \text{Var}_{\text{H}}\left[ \widehat{\phi} \right] \approx \frac{0.250\pm 0.001}{\lvert
    \alpha  \rvert^2 } + \frac{0.520\pm 0.010}{\lvert \alpha  \rvert^4}\, .
\end{equation}
Figure~\ref{fig:Excess}A summarizes the main result comparing the three
asymptotic variances for the canonical phase measurement (solid blue), MKII
(solid red), and non-Gaussian strategies (solid green and points).
Figure~\ref{fig:Excess}B, shows the excess Holevo variance compared to the
canonical phase measurement ($\text{Var}[ \cdot ] - \text{Var} [
\hat{\phi}_{\text{CPM}} ]$) for Heterodyne (purple), MKII (red), and
non-Gaussian strategies (green).

These non-Gaussian strategies implement a series of adaptive steps with
displacement operations optimized to maximize information gain, while ensuring
efficient phase estimators in the asymptotic limit. These strategies surpass the
best known Gaussian strategy in Eq. (\ref{eq:MKII_var}), and have the
same functional form as the canonical phase measurement in the asymptotic limit,
differing only by a scaling factor, thus providing the highest sensitivity among
physically-realizable measurements for single-shot phase estimation of coherent
states known to date.

Achieving this performance with non-Gaussian estimation strategies, however,
requires a deep understanding of the measurement process. To gain this
understanding, we use the mathematical framework of Bayesian optimal
experimental design, which provides a natural description of adaptive
non-Gaussian measurements. This allows us to optimize these strategies for
single-shot phase estimation with a Holevo variance given by Eq.
(\ref{eq:model}).

\subsection{\label{sec:BODE} Bayesian optimal design of experiments}

Phase estimation of coherent states based on photon counting with adaptive
coherent displacement operations can be defined in the context of Bayesian
optimal design of experiments. Optimal design of experiments allows for
improving statistical inferences about quantities of interest by appropriately
selecting the values of control variables known as \textit{designs}
\cite{Chaloner1995,Eli2016,suzuki2020quantumstate}. In this framework, it is
assumed that the experimental data $y$ (the measurement outcomes) can be modeled
as an element of the set
\begin{equation}
  \label{eq:models_1}
  \mathcal{P}_{\Phi} = \left\{ p(y \mid \mathbf{d}, \mathbf{\phi}), \, \, \, \mathbf{d} \in \mathcal{D} \, \, \, \mathbf{\phi} \in \Phi  \right\},
\end{equation}
where $\mathbf{d}$ is a parameter called design chosen from some set
$\mathcal{D}$ called design space, $\bm{\phi} \in \Phi$ is an unknown parameter
to be estimated, and the data $y$ comes from a sample space $\mathcal{Y}
\subseteq \mathbb{R}$. In this paradigm, the experimenter has full control over
the designs $\mathbf{d}$ and the ability to adjust them prior to making a
measurement. This allows for optimizing such measurement for estimating the
unknown parameter $\bm{\phi}$. Bayesian optimal design of experiments goes
beyond standard methods for parameter estimation based on Bayesian statistical
inference \cite{Morelli_2021,
  Mart_nez_Garc_a_2019,Berni2015,PhysRevA.85.043817,Oh_2014}, by providing the
suitable mathematical framework to ensure optimal designs to find efficient
estimators for a general parameter space \cite{Macieszczak_2014,
  Glatthard:2022izl, McMichael, Kleinegesse2019EfficientBE}.

In the Bayesian approach of optimal design \cite{Chaloner1995,Eli2016}, the
initial lack of knowledge about $\bm{\phi}$ is modeled as a prior probability
distribution $p(\bm{\phi})$. The measurement aims to reduce the uncertainty of
$\bm{\phi}$ as much as possible by the use of Bayes' theorem over the prior
distribution. In an estimation problem, the optimal choice for the designs $
\mathbf{d} \in\mathcal{D}$ maximize the expected value of a cost function
$U(\bm{d},\bm{\phi},y)$ with respect to the possible outcomes of $y$ and
$\bm{\phi}$:
\begin{equation}
  \begin{split}
    \mathbf{d}^{\text{opt}} &= \arg\max_{\mathbf{d} \in \mathcal{D}} \text{E} \left[ U(\mathbf{d},\bm{\phi},y) \right] \\
                     &= \arg\max_{\mathbf{d} \in \mathcal{D}} \int_{\mathcal{Y}} \int_{\Phi}
                       U(\mathbf{d},\bm{\phi},y) p(\bm{\phi} \mid \mathbf{d},y) d \bm{\phi} p(y \mid \mathbf{d}) dy\\
                     &= \arg\max_{\mathbf{d} \in \mathcal{D}} \int_{\mathcal{Y}} \int_{\Phi}
                       U(\mathbf{d},\bm{\phi},y)p \left( \bm{\phi}, y \mid \mathbf{d} \right) d \bm{\phi} dy \, .
   \end{split}
\end{equation}

A standard approach in optimal design of experiments considers choosing
$\mathbf{d}^{\text{opt}}$ at the beginning of the experiment an then sample data
from $p(y \mid \mathbf{d}^{\text{opt}}, \bm{\phi})$ for all subsequent trials.
An alternative approach considers dynamically updating $\mathbf{d}^{\text{opt}}$
on each trial, as more data is collected. The advantage of this approach is that
adaptive estimation strategies are never less efficient than the non-adaptive
ones \cite{Paninski2005}.

The implementation of Bayesian optimal design of experiments requires a cost
function, such as the Kullback-Lieber divergence between the prior and posterior
distributions \cite{Ver2000}:
\begin{equation}
  \label{eq:U-KL}
  \begin{split}
    U(\mathbf{d}, y) &= D_{\text{KL}}\left[p(\bm{\phi} \mid \mathbf{d},y) \mid \mid p(\bm{\phi}) \right] \\
                     &=   \int_{\Phi} p(\bm{\phi} \mid \mathbf{d},y) \log \left[ \frac{p(\bm{\phi} \mid \mathbf{d},y)}{p(\bm{\phi})} \right] d \bm{\phi}.
    \end{split}
\end{equation}
The Kullback-Lieber divergence provides a distance between probability
distribution $p(\bm{\phi} \mid \mathbf{d}, y)$ and $p(\bm{\phi})$
\cite{Cover2006}. If $p(\bm{\phi} \mid \mathbf{d}, y)$ is equal to
$p(\bm{\phi})$ then $U(\mathbf{d}, y)=0$ and there is not any gain of
information about $\bm{\phi}$ by measuring with design $\mathbf{d}$ and
outcome $y$.

Another cost function considered in optimal design is the conditional entropy between the plausible
values of $\bm{\phi}$ and $y$
\begin{equation}
  \label{eq:U-Entropy}
  \begin{split}
    U(\mathbf{d}) &= - H(\bm{\phi} \mid Y)\\
                  &= - \sum_{y \in \mathcal{Y}}p(y) \int_{\Phi} p(\bm{\phi} \mid \mathbf{d},y) \log \left[ p(\bm{\phi} \mid \mathbf{d}, y) \right] d \bm{\phi},
    \end{split}
  \end{equation}
  which is a measure of how much information is needed to describe the outcomes
  of the random variable $\bm{\phi}$ given that the value of another random variable
  $Y$ is known \cite{Cover2006}. However, we note that cost functions Eq.
  \eqref{eq:U-KL} and Eq. \eqref{eq:U-Entropy} can be related to the mutual
  information \cite{Cover2006}:
\begin{equation}
  \label{eq:U-MI}
  \begin{split}
    U(\mathbf{d}) = I(\bm{\phi} \mid Y) &= \text{E}_{y}\left[ D_{\text{KL}}\left[p(\bm{\phi} \mid \mathbf{d},y) \mid \mid p(\bm{\phi}) \right]   \right]\\
    &= H(p(\bm{\phi})) - H(\bm{\phi} \mid Y).
    \end{split}
\end{equation}
As a result, designs $d$ maximizing any of these cost functions in Eq.
\eqref{eq:U-KL}, Eq. \eqref{eq:U-Entropy}, or Eq. \eqref{eq:U-MI} are equivalent
\cite{Chaloner1995,Ryan2014,Eli2016}. Moreover, in the asymptotic limit,
maximizing these cost functions is equivalent to minimizing the determinant of
the covariance matrix (D-optimality design criteria), that is, in the
asymptotic limit, optimizing these cost functions is equivalent to optimizing
any member within the family of D-optimal designs \cite{Chaloner1995,Ver2000}.

Our goal is to apply the theory of Bayesian optimal design of experiments to the
problem of phase estimation of coherent states with photon counting and adaptive
coherent displacement operations. The adaptive non-Gaussian estimation strategy
consists of several parts: i) in the first adaptive step, it uses an specific
cost function and the prior information to choose the design; ii) then, it
performs a measurement; iii) based on the measurement outcome, it uses Bayes'
theorem to update the probability distribution; iv) and lastly, it uses a
recursive approach, where this posterior probability distribution becomes the
prior of the subsequent measurement step i). The estimation of $\bm{\phi}$
at each adaptive step is obtained from the maximum posterior estimator (MAP) of
the posterior probability distribution. This approach requires that the MAP
converge to the true value of the parameter when the number of measurements
increases.

In the adaptive mathematical framework of optimal experimental design, Paninski
\cite{Paninski2005} proved that under a set of regular modelling conditions
and in the case when $\phi \in \mathbb{R}$, cost functions
based on mutual information can allow for designs that lead to asymptotically
consistent and efficient MAP estimators with variance
\begin{equation}
  \label{eq:var_ad}
  \sigma^{2}_{\text{INFO}} = \left(  \argmax{C \in co\left( F(\phi; \mathbf{d})
      \right)}  \left \lvert  C  \right \rvert \right)^{-1}.
\end{equation}
Here $co\left( F(\phi; \mathbf{d}) \right)$ denotes the convex closure of the
set of Fisher information functions $F(\phi; \mathbf{d})$. In other words, the
estimations produced by $\widehat{\phi}$ converge to $\phi$ (consistency), and
the distribution of $\widehat{\phi}$ tends to a normal distribution with mean
$\phi$ and variance given by Eq. (\ref{eq:var_ad}) when the number of adaptive
steps tends to infinity (efficiency).

Formally, the regularity conditions introduced in \cite{Paninski2005} can be stated as follows:
\begin{enumerate} \label{RC1} \item  The parameter space $\Phi$ is a compact metric space.
\item The log likelihood, $\log \left( p(y \mid \phi, \mathbf{d}) \right)$ is
  uniformly Lipschitz in $\phi$ with respect to some dominating measure on
  $\mathcal{Y}$.
\item The likelihood function is identifiable for $\phi$, that is, the
  likelihood function has a unique global maximum.
\item The prior distribution, assigns a positive probability to any neighborhood
  of the real value of $\phi$.
\item The Fisher information functions $F(\phi; \mathbf{d})$ are well defined
  for any $\phi \in \Phi$ and $\mathbf{d} \in \mathcal{D}$.
\item The maximum of the convex closure of the set of Fisher information
  functions $\left\{ F(\phi, \mathbf{d}) \mid \mathbf{d} \in \mathcal{D}, \,
    \phi \in \Phi \right\}$ must be positive-definite, i.e., $\max_{C \in
    co\left( F(\phi; \mathbf{d}) \right)} \left \lvert C \right \rvert >0$.
\end{enumerate}

We note that in the case of estimation of a scalar parameter, the maximization of
mutual information is equivalent to the minimization of the mean square error
(MSE) \cite{Paninski2005}:
\begin{equation}
  \label{eq:mse}
  \begin{split}
    \rm{MSE}(\widehat{\phi}) &= \rm{E} \left[  \right( \widehat{\phi} - \phi \left)^2 \right] \\
                             &=\text{Var}\left( \widehat{\phi} \right) + \left( \rm{E}\left[ \widehat{\phi} \right] - \phi \right)^2.
  \end{split}
\end{equation}

This shows that the MSE is a trade-off between the estimator's variance and its
bias. As a result, since the phase is a scalar quantity, in the asymptotic limit
both the mutual information Eq. (\ref{eq:U-MI}) and the mean square error Eq.
(\ref{eq:mse}) are appropriate cost functions to find optimal estimation
strategies. Even more, for unbiased estimators (such as the MAP estimator on the
asymptotic limit), the MSE corresponds to the estimator variance, then the
maximization of mutual information is equivalent to the minimization of
estimator variance. In practice, however, an estimation strategy would use a
cost function that can be calculated efficiently and with a high rate of
convergence.
\\
\\
\textbf{\textit{{\label{sec:Phase_Estimation}Phase estimation
        in coherent states.}}}
\\
Optimal phase estimation of coherent states of light aims to obtain the best
estimate, from the outcomes of a physical measurement, of an unknown phase $\phi
\in [0, 2\pi)$ encoded in a coherent state $\rho(\phi) = \ket{e^{i \phi }
  \alpha}\bra{e^{-i \phi }\alpha} \,$. The most general description of a
physical measurement is given by a POVM, a positive operator-valued measure. A
measurement $M$ with a discrete set of outcomes $\left\{ m \, | \, m \in S
  \subseteq \mathbb{Z} \right\}$, can be represented as a POVM $M = \left\{ M(m)
  \, | \, m \in S \right\}$, where the operators $M(m)$ are positive bounded $
M(m) > 0$ and resolve the identity $\sum_{m} M(m) = I$, ${\forall m \in S}$
\cite{Nielsen2000}. By the Born rule, the probability for $m$ conditioned to
$\phi$ is
\begin{equation}
  \label{eq:BR}
 \text{ Tr}\left[ M (m)\rho(\phi) \right] =
  p\left( m \mid \phi \right)\, .
\end{equation}

According to information theory, if an estimator $\widehat{\phi}$ of ${\phi}$ is
constructed using a sample from a POVM $M$, the limit in the accuracy of
$\widehat{\phi}$ is given by the Crámer-Rao Bound \cite{Braunstein1994,Paris2009}
\begin{equation}
  \label{eq:CRB}
 \text{Var}_{\phi}\left[ \widehat{\phi} \right] \geq \frac{1}{F_{M}(\phi)}.
\end{equation}
Here $F_{M}(\phi)$ is the Fisher information of $M$ about $\phi$, which quantifies
how much information about $\phi$ is carried in a sample from $M$:
\begin{equation}
  \label{eq:Fisher_Inf}
  F_{M}(\phi) = \text{E}_{\phi} \left[  \left(  \frac{\partial}{\partial \phi} \left[ \log\left( p\left( m \mid \phi \right) \right) \right]  \right)^2  \right].
\end{equation}

Since the Fisher information in Eq. (\ref{eq:Fisher_Inf}) depends on the POVM
$M$, the maximization of the Fisher information over all POVMs provides the
lowest possible Cramér-Rao bound. This maximum Fisher information over all POVMs
is known as the quantum Fisher information $F_{\text{Q}}$ (QFI), and the lowest possible
bound in the accuracy of $\widehat{\phi}$ is known as the quantum Cramér-Rao bound
(QCRB) \cite{Helstrom1969,Holevo1973,Paris2009}. In the case of phase estimation
of coherent states $F_{\text{Q}} = 4 \lvert\alpha \rvert^2$, and the QCRB is:
\begin{equation}
  \label{eq:FQ_Coherent}
  \text{Var}_{\phi}\left[ \widehat{\phi} \right] \geq \frac{1}{4 \lvert \alpha  \rvert^2}.
\end{equation}

In the limit of large photon number $\lvert \alpha \rvert \gg 1$, the QCRB is
saturated by the canonical phase measurement (the optimal phase measurement),
which is described by the POVM \cite{Holevo2011}:
\begin{equation}
  \label{eq:canonical}
  M(\widehat{\phi}) = \frac{1}{2 \pi} \sum_{n,m= 0}^{\infty} \ket{n}\bra{m}e^{\const{i} \left( n-m \right) \widehat{\phi}},
\end{equation}
where $\ket{n}$ is an eigenstate of the number operator $\hat{n}$. The operator
$M(\widehat{\phi})$ is an element of the canonical phase measurement whose outcome
is a number $\widehat{\phi} \in \left[ 0, 2\pi \right)$, which can be used as an estimation for  $\phi$.

The optimality of the canonical phase measurement indicates that its Holevo
variance
\begin{equation}
  \label{eq:Var_can}
  V_{\text{CPM}} = \left  \lvert e^{-\alpha^2}
    \sum_{n=0}^{\infty}\frac{\alpha^{2n+1}}{n!\sqrt{n+1}} \right
  \rvert^2 -1\, ,
\end{equation}
is the fundamental bound of estimation precision \cite{Holevo2011,
  Wiseman1995}. Although there are proposals that attempt to implement
this POVM, they have not been able to reach the fundamental bound
Eq.~(\ref{eq:Var_can}), or  Eq.~(\ref{eq:CPM_var}). For instance, in \cite{Juha2013} it was possible to obtain the canonical measurement distribution as the marginal of a
joint measurement in phase space producing a worse performance in the
context of phase estimation. Moreover, the canonical phase measurement was implemented for the case of one-photon wave packet using quantum feedback \cite{Martin2019}. However, for the case of higher dimensional states, such as coherent states, this problem remains open.

While there is not a satisfactory known method to implement the canonical phase
measurement, Gaussian strategies serve as a standard of physically realizable
measurement techniques for phase estimation of coherent states. The natural
benchmark in the Gaussian strategies is the heterodyne detection, whose variance
is lower bounded by $\text{Var}\left[ \widehat{\phi}_{\text{Het}} \right]=1/2
\lvert \alpha \rvert^2$ \cite{wiseman98}. Several adaptive Gaussian schemes have
been shown to exceed the lower bound for heterodyne detection. The most
efficient Gaussian measurement reported to date, termed the Adaptive Mark-II
(MkII) strategy \cite{Wiseman1995}, has a variance in the limit of $\lvert
\alpha \rvert \gg 1$ given by Eq.~(\ref{eq:MKII_var}). Nonetheless, these
adaptive Gaussian strategies cannot reach the performance for the canonical
phase measurement in Eq.~(\ref{eq:CPM_var}) \cite{wiseman98}.

The proposed non-Gaussian strategies for single-shot phase estimation of
coherent states are based on optimized adaptive measurements with photon number
resolving detection. These non-Gaussian strategies are able to outperform the
best known Gaussian strategies and closely follow the performance of the
canonical phase measurement in the limit of large photon number.

\subsection{\label{sec:Adaptive}Adaptive non-Gaussian phase estimation}

The proposed non-Gaussian adaptive estimation strategies based on adaptive
photon counting \cite{DiMario2020} aim to estimate the phase $\phi \in \Phi =
[0,2 \pi)$ of a coherent state $\rho(\phi) = \ket{e^{\const{i} \phi }
  \alpha}\bra{e^{-\const{i} \phi }\alpha} \,$ with mean photon number
$\text{E}\left[ \hat{n} \right] = \lvert {\alpha} \rvert^2$ using a finite
number of adaptive measurement steps, and based on the prior information
$p(\phi)$ about $\phi$. In every adaptive step $l=1,2,\cdots,L$, the input
coherent state with energy $\lvert {\alpha} \rvert^2 /L$ interferes with a local
oscillator field, which implements a displacement operation $\hat{D}\left( \beta
\right)\ket{\alpha} = \ket{\alpha + \beta}, \, \beta \in \mathbb{C},$ with phase
and amplitude chosen by some rule, in general depending on previous measurement
outcomes. This is followed by a photon number detection measurement with a given
photon number resolution (PNR) $m$ of the detectors \cite{becerra15}, $m \in
\mathbb{N}$. In practice, since the energy in each adaptive step
  is $\lvert {\alpha}\rvert^2 /L$, the strategy will only require moderate PNR
  resolution ($m<10$) as $L$ increases. 

In the first adaptive measurement $l=1$ \cite{DiMario2020}, the strategy makes a
random guess hypothesis $\beta_0 \in \mathbb{C}$ about the optimal $\beta$, and
applies the POVM {\small{
\begin{eqnarray}
  \label{eq:POVM_1}
\left\{
  \hat{D}(\beta_0)\ket{n}\bra{n}\hat{D}^{\dagger}(\beta_0) \right\}_{n=0}^{m-1} \cup \nonumber \\
  \left\{
  \mathbb{I} - \sum_{n=0}^{m-1} \hat{D}(\beta_0)\ket{n}\bra{n}\hat{D}^{\dagger}(\beta_0)
\right\}
\end{eqnarray}
}} over the state $\ket{e^{\const{i}\phi} \alpha/\sqrt{L}}$. In the POVM in
Eq. (\ref{eq:POVM_1}), the sum over Fock states $\ket{n}\bra{n}$
models the photon detection on the displaced state with a detector
with PNR($m$) \cite{becerra15}. 
 The corresponding
	Wigner function describing a photon-number resolving detector shows non-Gaussian features with negative values. For this
	reason, these adaptive estimation techniques are called
	“non-Gaussian”, despite that the estimator produced is
	asymptotically normal (this result will be proved in the remainder of this
	section).

Given the detection outcome $n_1$ in $l=1$, the posterior probability distribution becomes
\begin{equation}
  \label{eq:post_1}
  p\left( \phi \mid n_1; \beta_0 \right) \propto  \mathcal{L}(\phi \mid
  n_1; \beta_0)p(\phi),
\end{equation}
where $\mathcal{L}\left( \phi \mid n_1; \beta_0 \right)$ is the likelihood
function given by
\begin{equation}
  \begin{split}
    \mathcal{L}\left( \phi \mid n_1; \beta_0 \right) &= p\left( n_1
      \mid \phi; \beta_0 \right) \\ &= \operatorname{Tr}\left[
      \hat{D}(\beta_0)\ket{n_1}\bra{n_1}\hat{D}^{\dagger}(\beta_0)
      \rho(\phi) \right]\, .
    \end{split}
\end{equation}
The phase estimate $\phi_{1}$ in this adaptive step corresponds to the MAP
estimator $\widehat{\phi}_{\text{MAP}}$, $\phi_1 = \widehat{\phi}_{\text{MAP}}(n_1)$, with the
posterior probability distribution in Eq. \eqref{eq:post_1}. Using the posterior
phase distribution in Eq. (\ref{eq:post_1}) as the prior for the next adaptive
step $l=2$, the strategy optimizes a cost function $U(\beta)$ to obtain the next
value of $\beta$ denoted as $\beta_{1}$, and implements the POVM in Eq.
\eqref{eq:POVM_1} with $\beta_1$. The Bayesian updating procedure is repeated at
each step $l \geq 2$. After $l$ adaptive measurements the posterior probability
distribution becomes
\begin{equation}
  \label{eq:post}
  \begin{split}
  p(\phi \mid \mathbf{n}, \bm{\beta}) &= p(\phi \mid n_{l},...,n_{1},
  \beta_{l-1},...,\beta_{0})\\ &\propto \prod_{i=1}^{l} p(n_i \mid \phi,
  \beta_{i-1})p(\phi).
  \end{split}
\end{equation}
Here $n_i$ is the observed photon detection at step $i$. Using the MAP on this
phase distribution, we obtain the $lth$ estimation $\widehat{\phi}_l$. The procedure
is repeated until the last measurement step $L$.

This parameter estimation strategy is a particular case of Bayesian optimal
design of experiments, where the parameters $\beta \in \mathbb{C}$ are the
designs, and which are optimized to estimate a phase $\phi \in [0, 2\pi)$. Since
the optimal value for $\beta$ on each adaptive step depends on all previous
detection results, the cost function to be optimized is a function of the
posterior distribution in Eq. (\ref{eq:post}). A suitable choice of cost
function, such as the mutual information or the estimator variance, can provide
a sequence of estimations $\widehat{\phi}_n$ that approaches the true value of
$\phi$ \cite{DiMario2020}.

In the case of estimation of cyclic parameters, such as phase estimation, the
posterior distribution in Eq.~\eqref{eq:post} is $2 \pi$ periodic, and the
moments of $\widehat{\phi}$ cannot be calculated as in the linear case
\cite{Holevo2011}. In such situations, the first moment of the cyclic random
variable $X$ is defined as $\rm{E}\left[ e^{\const{i} X} \right]$, and the dispersion of an
estimator $\widehat{\phi}$ is calculated using the Holevo Variance
\cite{Holevo2011}:
\begin{equation}
  \label{eq:Holevo_Variance}
  \text{Var}_{\text{H}}\left[ \widehat{\phi} \right] = \frac{1}{ \left \lvert \rm{E}
      \left[ e^{\const{i} \widehat{\phi}} \right]  \right \rvert^2   } -1\, ,
\end{equation}
which is the analogous to the mean square error. The minimization of the
uncertainty about $\phi$ (positive square root of Eq.
(\ref{eq:Holevo_Variance})), requires maximization of $S(\beta, m)=\left \lvert
  \rm{E} \left[ e^{\const{i} \widehat{\phi}} \right] \right \rvert$, known as the sharpness
of the distribution. Then, the suitable cost function for the adaptive protocol
is the average sharpness:
\begin{equation}
  \label{eq:av_sharp}
  \bar{S}(\beta, m) = \sum_{n=0}^{m} p(n) \left \lvert  \int_{\Phi}e^{\const{i} \phi} p\left( \phi \mid n, \beta \right) d \phi \right   \rvert.
\end{equation}
\\
\\
\textbf{\textit{{\label{sec:IdentifiabilityLkhd}Identifiability of Likelihood.}}}
\\
To guarantee a consistent asymptotic estimator the optimized estimation
strategies require to satisfy the regularity conditions  1-6 described in Sec. \ref{RC1}. For phase
estimation, the regularity conditions $1 \text{ and } 2$ are satisfied given
that $\phi$ is an interior point of $\Phi = [0, 2 \pi)$. Moreover, given that
the probability
{\small{
\begin{equation}
  \label{eq:prob}
  \begin{split}
    &p(n \mid \phi; \beta) = \text{Tr}\left[ \ket{ \frac{ \alpha e^{\const{i} \phi}}{\sqrt{L}} }\bra{  \frac{  \alpha e^{\const{i} \phi}}{\sqrt{L}} } \hat{D}(\beta)\ket{n}\bra{n}\hat{D}^{\dagger}(\beta)   \right]  \\
    &= \frac{\exp \left( - \lvert \frac{ \alpha e^{\const{i} \phi}}{\sqrt{L}} - \beta
        \rvert^2 \right) \lvert \frac{ \alpha e^{\const{i} \phi}}{\sqrt{L}} - \beta
      \rvert^{2n} }{n!}, \, \alpha \in \mathbb{R}_{+}, \, \beta \in \mathbb{C}
  \end{split}
\end{equation}
}}is well defined and two times differentiable, the conditions $4$, $5$, and
$6$ are directly satisfied. However, the condition $3$ is not satisfied in
general. If one chooses $\beta$ as the value that maximizes the mutual
information (\ref{eq:U-MI}) or the average sharpness (\ref{eq:av_sharp}), the
resulting likelihood function can have in general two maxima, that is, a
non-identifiable likelihood function \cite{Huang2017}. In that case it is not
possible to guarantee the existence of a consistent estimator.

To address the challenge of working with non-identifiable likelihood functions,
we use designs with a fixed relation between the phase of $\beta$ and the
amplitude $|\beta|$, given by $\beta = f(\theta) e^{\const{i} \theta}$, with $\theta \in
[0, 2\pi)$ and $f(\theta)$ a real function. These experimental designs result in
a cost function $U$ that is a function of $\theta$.

To see how this method solves the problem of non-identifiability,
we observe that the probability $p(n \mid \phi; \beta) $ in Eq. (\ref{eq:prob})
is Poisson distributed,
\begin{equation}
  \label{eq:prob_2}
  \begin{split}
    p(n \mid \phi; \beta = \lvert \beta \rvert e^{\const{i} \theta} ) &=
    \frac{e^{- \lambda} \cdot \lambda^n}{n!},
  \end{split}
\end{equation}
with $\lambda = \lvert \alpha \rvert^2/L + \lvert \beta \rvert^2 - 2 \lvert
\alpha \rvert \lvert \beta \rvert \cos\left( \phi- \theta \right)/\sqrt{L}$. Then, for
$L$ adaptive steps with results $\mathbf{n} = (n_1,...,n_L)$, the likelihood
function is the product of $L$ probability functions of the form of Eq.
(\ref{eq:prob_2}):
\begin{equation}
  \label{eq:prob_2b}
  \begin{split}
    \mathcal{L}(\mathbf{n},\phi)=\prod_{i=1}^L p(n_i \mid \phi; \beta_i =
    \lvert \beta \rvert e^{\const{i} \theta_i} ) &=\prod_{i=1}^L \frac{e^{- \lambda_i}
      \cdot \lambda_i^{n_i}}{n_i!}\, .
  \end{split}
\end{equation}
Here, each $\theta_i$ depends on the cost function and previous POVM outcomes.
The choice of the experimental designs with $|\beta|=f(\theta)$ can force the
adaptive strategy to change $\theta_i$ in each step. In this case, the
likelihood function in Eq. \eqref{eq:prob_2b} becomes identifiable because the
product of probability functions with different $\theta_i$ produces a likelihood
with a global maximum. To see this, note that if $\theta_i=\theta$ is fixed,
even if the parameter $|\beta|$ is different in each adaptive step of the
protocol, the likelihood function from a sequence of independent random
variables with probability function given by Eq. (\ref{eq:prob_2}) has two
maxima over $\Phi$. One of the maxima will always be around $\phi$ and the other
will depend on the value of $\theta$. On the other hand, if the strategy allows
$\theta$ to change at each adaptive step, the second maximum is suppressed,
since the functions whose product constitutes the likelihood Eq.
\eqref{eq:prob_2b} will each have a second maxima at different positions
$\theta_i \neq \theta_j \, (i \neq j)$. As a result, with the experimental
designs $\beta=f(\theta)\exp(\const{i}\theta)$, the likelihood functions satisfy all the
regularity conditions described in Sec. \ref{RC1}.
\\
\\
\textbf{\textit{{\label{sec:adabtivedesign}Bayesian optimal design of $|\beta|$.}}}
\\
Any viable estimation strategy should aim to achieve the QCRB (\ref{eq:FQ_Coherent}). Therefore,
natural choice for $\lvert {\beta} \rvert$ is the one that maximizes the
Fisher information. For a discrete random variable, the Fisher information is
given by \cite{Paris2009,Escher2011}:
\begin{equation}
  \label{eq:FI_discrete}
  F(\phi) = \sum_{n=0}^{\infty} p(n \mid \phi) \left( \frac{\partial}{ \partial \phi} \log \left( p(n \mid \phi ) \right)    \right)^2.
\end{equation}
The Fisher information for a particular design $\beta = \lvert {\beta } \rvert
e^{\const{i} \theta}$ for Poisson distributions Eq.~(\ref{eq:prob_2}) results in
\begin{equation}
  \label{eq:FI_2}
  F(\phi; \beta) = \frac{4 \lvert \alpha \rvert^2 \lvert \beta \sin^2(\phi-
\theta) \rvert^2 }{ \lvert \alpha \rvert^2 + L\lvert \beta \rvert^2 -2 \lvert
\alpha \rvert \lvert \beta \rvert \cos(\phi- \theta) \sqrt{L} }.
\end{equation}
Optimizing over $|\beta|$, the Fisher information becomes:
\begin{equation}
  \label{eq:FI_opt}
  F(\phi, \beta_{\text{opt}}) = 4 \lvert \alpha  \rvert^2/L,
\end{equation}
where
\begin{equation}
  \label{eq:beta_opt}
  \beta_{\text{opt}}= \frac{\lvert \alpha
    \rvert}{\sqrt{L}\cos(\phi- \theta)}
\end{equation}
is the value of $|\beta|$ that maximizes the Fisher information. Unfortunately, since $\beta_{\text{opt}}$ depends on $\phi$, its implementation is not
practical, because it would be required to already know a priori the very same
parameter that one wants to estimate. To address this problem, the optimal
Bayesian design $\beta_{\text{opt}}$ can be estimated as:
\begin{equation}
  \label{eq:beta_opt_hat}
  \widehat{\beta}_{\text{opt}}= \frac{\lvert \alpha
    \rvert}{\sqrt{L}\cos(\widehat{\phi}- \theta)},
\end{equation}
where $\widehat{\phi}$ is the MAP estimator. As a result, the non-Gaussian estimation
strategy has now only one design, the phase $\theta$. With $\beta=\widehat{\beta}_{\text{opt}}$, the  Fisher information becomes:
{\small{
\begin{equation}
  \label{eq:fishernotperfect}
F(\Delta, \widehat{\beta}_{\text{opt}})=\frac{4\,\sin ^2(\Delta)\,\alpha^2}{L(\cos ^2\left(\delta+\Delta
 \right)-2\,\cos (\Delta)\,\cos \left(\delta+\Delta\right)+1)}\, ,
\end{equation}
}}
where $\delta=\widehat{\phi}-\phi$ and $\Delta=\phi-\theta$.

Note that $F(\Delta, \widehat{\beta}_{\text{opt}})$ becomes a random variable
with outcomes depending on $\widehat\phi$ through
$\widehat{\beta}_{\text{opt}}$. Moreover, for a random initial design
$\theta_1$, a cost function given by the expected sharpness or the mutual
information with $\widehat{\beta}_{\text{opt}}$ in Eq.~\eqref{eq:beta_opt_hat},
the likelihood function in Eq.~\eqref{eq:prob_2b} becomes identifiable. As a
result, the non-Gaussian strategy then leads to consistent and efficient MAP
estimators \cite{Paninski2005}.
\\
\\
\textbf{\textit{{Asymptotic behavior.}}}
\\
In general $\delta=\widehat{\phi}-\phi \neq 0$, and the Fisher information has a
strong dependence on $\theta$. For example, if $\theta\to\phi$ then $F(\Delta=0,
\widehat{\beta}_{\text{opt}})\to 0$, and negligible information is gained when
the system is measured. Therefore, the optimal value of $\theta$ should result
in the value of $\Delta$ that maximizes the expected value of $F(\Delta,
\widehat{\beta}_{\text{opt}})$. By observing that the expected value
$\rm{E}[\delta^{2n+1}]=0$, with $n\in\mathbb{N}$, we see that the optimal value
of $\Delta$ is $\pi/2$. As a result, an efficient adaptive strategy would make
$\Delta_L=\phi-\theta_L$ tend to $\pi/2$ as $L$ increases. However, in the limit
of $\Delta\to\pi/2$ and $\delta\to 0$, $|\widehat{\beta}_{\text{opt}}|\to\infty$
(see Eq. (\ref{eq:beta_opt_hat})), and we expect that when the strategy is
implemented $\Delta<\pi/2$. These findings are consistent with our numerical
simulations of the non-Gaussian adaptive strategy, where we observe that for $L
\gg 1$, $\Delta\to\pi/2-\epsilon$, with $\epsilon$ a small positive real number,
and $|\beta|$ stays finite around a value that the PNR detector in the strategy
can resolve. Note that $\hat\beta_{\text{opt}}$, which maximizes the Fisher
information given $\theta$, does not necessarily maximizes the cost function
$U(\beta)$ for $\beta\in \mathbb{C}$. However restricting $\beta$ to the set
that maximizes the Fisher information makes the phase of $\beta$ change in each
step forcing the likelihood to be identifiable. Moreover, when $\hat\phi \to
\phi$, $\hat\beta_{\text{opt}}$ tends to the value that maximizes $U$
\cite{DiMario2020}.

Given that the Fisher Information is additive, for any $L$ measurements made
using the optimal design, $\beta_{\text{opt}}$ in Eq. (\ref{eq:beta_opt}), the total
Fisher information for this design corresponds to the quantum Fisher information
$4|\alpha|^2$. However, since $\beta_{\text{opt}}$ is not known, it is not possible to
choose a design $\beta$ for which its Fisher information equals the quantum
Fisher information. This is already implied by the fact that the canonical phase
measurement does not saturate the Cramér-Rao bound. To show this, we observe
that for and estimator very close to the true value $\delta=\widehat{\phi}-\phi\ll 1$
for $L$ adaptive measurements, the Fisher information
(Eq.~\eqref{eq:fishernotperfect}) is
\begin{equation}
  \label{eq:fishernotperfect2}
F(\phi, \widehat{\beta}_{\text{opt}})\approx 4\,|\alpha|^2(1-\delta^2)\, .
\end{equation}
On the other hand, the best possible estimator of $\delta$ in each step
satisfies $E\left[ \delta^2\right]\geq 1/4|\alpha|^2$, so that
\begin{equation}
F(\phi, \widehat{\beta}_{\text{opt}})\lesssim 4\,|\alpha|^2-1\, ,
\end{equation}
independently of $L$. We conclude that the adaptive non-Gaussian strategies do
not saturate the Cramér-Rao bound for finite $|\alpha|$. Nevertheless, these
adaptive estimation schemes outperform the most sensitive Gaussian strategy
known to date, and show a similar asymptotic scaling as the canonical phase
measurement.
\\
\\
\subsection{\label{sec:Performance}Performance of Non-Gaussian Adaptive Strategies}

We numerically investigate the performance of the estimator
produced by non-Gaussian adaptive strategies for phase estimation for
different number of adaptive steps $L$, photon number resolution PNR($m$), and
average photon number $|\alpha|^2$. To assess the performance of these
strategies, we compare our results with the best known Gaussian measurement for
phase estimation, termed Mark II (MKII) strategy \cite{wiseman98}, and with the
canonical phase measurement. As discussed in Sec. \ref{sec:BODE}, the
performance of non-Gaussian adaptive strategies using cost functions including
the Kullback-Lieber divergence Eq. \eqref{eq:U-KL}, conditional entropy Eq.
\eqref{eq:U-Entropy}, mutual information Eq. \eqref{eq:U-MI}, and expected
sharpness Eq. \eqref{eq:av_sharp} are equivalent in the asymptotic limit. In our
numerical simulations, however, we use the expected sharpness in Eq.
\eqref{eq:av_sharp} as the cost function for the optimization of the strategy,
because it substantially reduces the number of operations for this optimization
and the computational overhead.

\begin{figure*}[htbp!]
\centering
  \includegraphics[width=\textwidth]{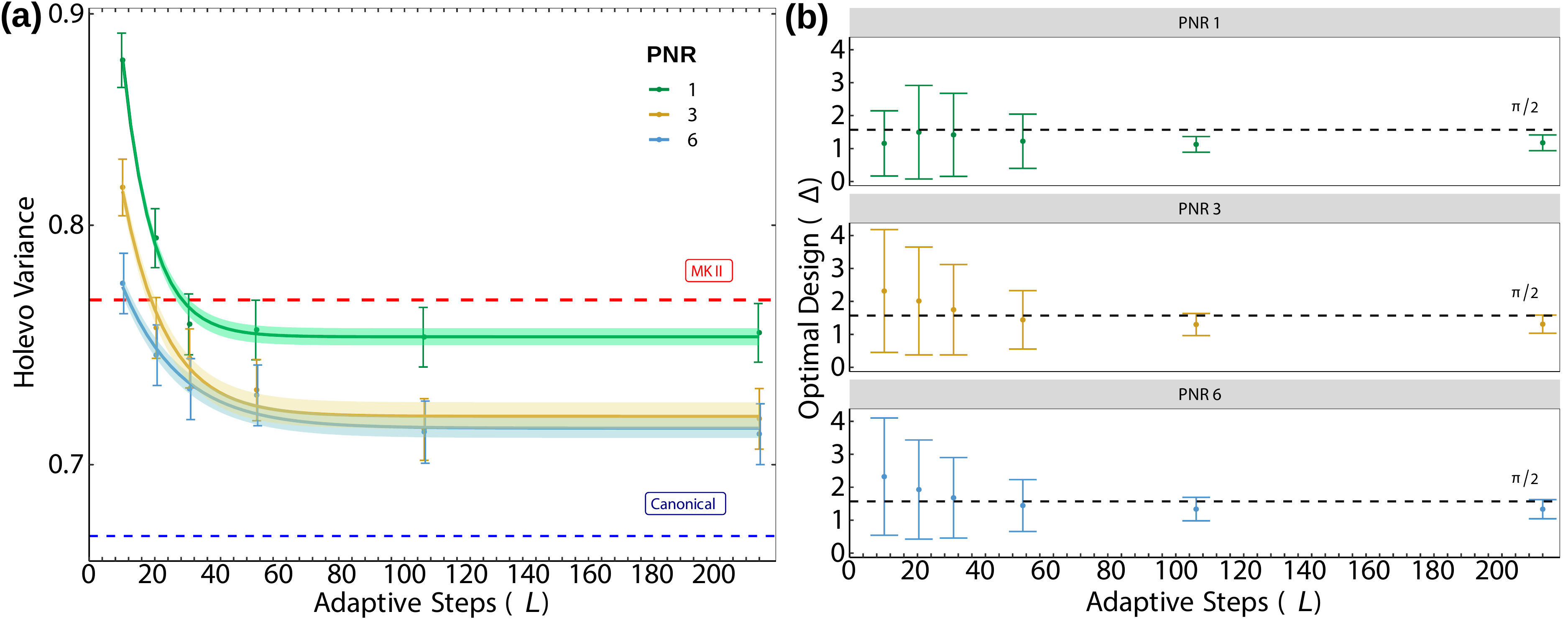}\quad
  \caption{\textbf{Asymptotic limit for the estimator of Holevo variance and
      optimal design}. \textbf{a}: Holevo variance for $\alpha =1$ and PNR $m=1,
    3, 6$ as a function of adaptive steps $L$. Note that the non-Gaussian
    strategy surpasses the MKII strategy (red dashed line at $y=0.767$) with $L
    > 100$, $L > 30$ and $L> 20$ for PNR($1$), PNR($3$) and PNR($6$),
    respectively. The lines are obtained fitting the exponential model Eq.
    (\ref{eq:fit_expo}). Estimates for these strategies result in
    $(A,B,C,\mathrm{RSE})=( 0.11 \pm 0.02,0.059 \pm 0.014 ,0.7145 \pm
    0.003,0.00058)$ for PNR($6$), $(A,B,C,\mathrm{RSE})=(0.22 \pm 0.04,0.082 \pm
    0.0157,0.719 \pm 0.004,0.00076)$ for PNR($3$), $(A,B,C,\mathrm{RSE})=(0.41
    \pm 0.05,B= 0.117 \pm 0.013, 0.7517 \pm 0.0027,0.00052)$ for PNR($1$). The
    shaded region represents one standard deviation. \textbf{b}: Optimal design
    as a function of $L$. As $L$ increases the optimal design tends to $\pi/2$.
    Numerical data are obtained with $10000$ Monte Carlo sequences. Error bars
    represent a 1-$\sigma$ standard deviation.}
  \label{fig:LowMPN}
\end{figure*}

In our numerical analysis, we use Monte Carlo simulations and generate
sufficient numerical data samples to reduce statistical uncertainties.
Fig.~\ref{fig:LowMPN}A shows the Holevo variance for the non-Gaussian adaptive
strategy, as a function of the number of adaptive steps $L$ for $|\alpha|^2=1$,
for different PNR: $m=1,3,6$. We observe that the adaptive non-Gaussian scheme
with PNR($1$) and $L \geq 100$ (green dots with error bars) outperforms the most
sensitive Gaussian strategy known to date, the MKII, (red dashed line).
Moreover, strategies with PNR($3$) (yellow dots with error bars) and PNR($6$)
(light blue dots with error bars) outperform the MKII strategy with only
$L\approx 30$ and $L \approx 20$, respectively, achieving a smaller Holevo
variance with fewer adaptive measurements compared to PNR(1). We have observed
similar behavior for non-Gaussian strategies optimizing different cost
functions, such as the mutual information.

To investigate the asymptotic behavior of the adaptive non-Gaussian strategy, we
assume an exponential dependence for the Holevo variance with $L$ (solid lines
in Fig.~\ref{fig:LowMPN}A):
\begin{equation}
  \label{eq:fit_expo}
y(L,\alpha) =  \text{A} e^{-\text{B} \cdot L}+\text{C}.
\end{equation}

The exponential model is a few parameter model that allows us to
  quantify asymptotic trends when the datasets have rapidly decaying tails.
  This is a widely used model for studying the asymptotic
    scaling of estimators as a function of resources in diverse metrological
  problems \cite{Glatthard:2022izl,Wang2021,PhysRevX.2.041006}.

We fit the numerical data from Monte Carlo simulations to Eq.
(\ref{eq:fit_expo}) to estimate the constants $A$, $B$, and $C$. Our results
show that the asymptotic constant $C=0.751 \pm 0.002$ for PNR(1), $C=0.719 \pm
0.004$ for PNR(3), and $C=0.714 \pm 0.003$ for PNR(6). We observe that these
values are smaller than the asymptote of the MKII Gaussian strategy, but larger
than the one for the canonical phase measurement ($0.673$, blue dashed line). We
note that while $C$ for PNR(3) is larger than for PNR(6), they are statistically
equal due to their uncertainties. This prevents us from drawing any
conclusions for larger values of $m$.

Figure~\ref{fig:LowMPN}\textbf{b} shows the design
$\Delta=\widehat{\phi}-\theta$, the phase of the displacement field, as a
function of $L$. We observe that for non-Gaussian adaptive strategies with
increasing PNR, $\Delta$ tends to $\pi/2$ for large $L$ ($L=200$). This
observation is consistent with the theoretical framework of optimal design of
experiments (Sec.~\ref{sec:adabtivedesign}), which states that
$\Delta_{optimal}=\pi/2$ for $L\to\infty$. Moreover, as PNR increases, the
strategies show a faster convergence to the asymptotic value of the Holevo
variance, which translates in a smaller error in the estimation (see
Fig.~\ref{fig:LowMPN}\textbf{a}.) These observations further support our
theoretical model of non-Gaussian strategies for phase estimation.

\begin{figure}[htbp!]
  \includegraphics[scale=0.44]{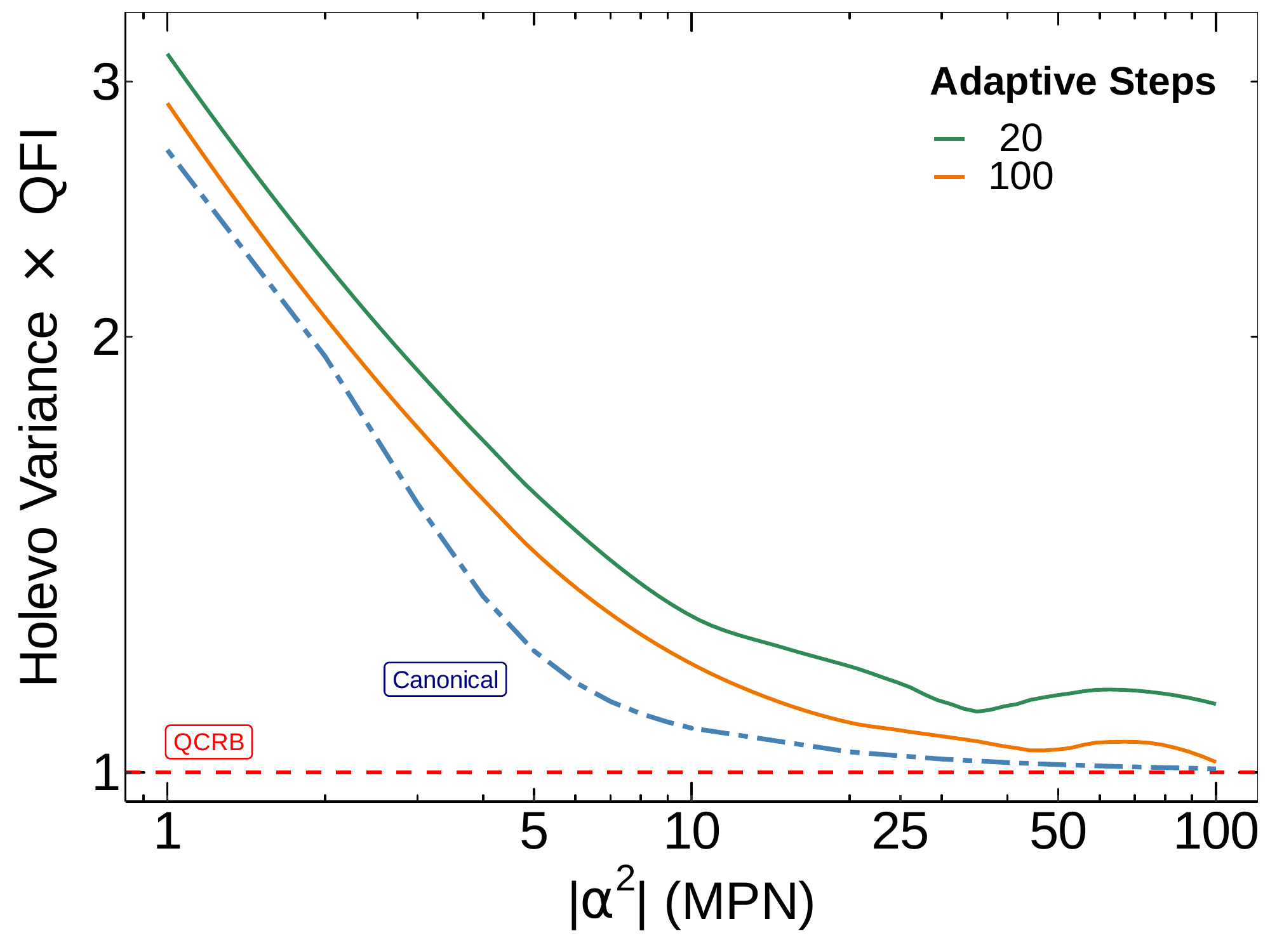}
  \caption{ \textbf{Estimator variance produced by the non-Gaussian strategy}.
    Holevo variance of adaptive Non-Gaussian strategies as a function of $\lvert
    \alpha \rvert^2$, normalized to QCRB, for different adaptive measurement
    steps (solid lines), together with the canonical phase measurement (blue
    dashed line). For $L=100$ the Holevo variance differs from the QCRB by $3\%$
    exemplifying the tendency to the ultimate precision limit in the regime of
    large number of photons in the asymptotic limit. Simulations consist of
    $10000$ Monte Carlo simulation sequences with PNR $m =3$. }
  \label{fig:Asymp}
\end{figure}

We investigate the asymptotic performance of the estimator
  variance produced by the non-Gaussian strategy for large $|\alpha|^2$. Figure
~\ref{fig:Asymp} shows the Holevo variance for the non-Gaussian adaptive
strategy as a function of $|\alpha|^2$ for different $L$ normalized to the QCRB.
We observe that for large $|\alpha|^2$ (with $L=100$), the adaptive non-Gaussian
strategy tends to the CRLB.

To build a model for the performance of this strategy for large $|\alpha|^2$, we
propose three candidate models for the Holevo variance for $L \gg 1$:

\begin{subequations} \label{eq:models}
  \begin{eqnarray}
    \widehat{y}_1 &=& \frac{A_1}{\lvert \alpha \rvert^2} + \frac{A_2}{\lvert \alpha
      \rvert^3} + \frac{A_3}{\lvert \alpha \rvert^4}, \label{eq:modelsa}\\
    \widehat{y}_2 &=& \frac{A_1}{\lvert \alpha \rvert^2} + \frac{A_2}{\lvert \alpha
      \rvert^3}, \label{eq:modelsb}\\
    \widehat{y}_3 &=& \frac{A_1}{\lvert \alpha \rvert^2} + \frac{A_3}{\lvert \alpha
      \rvert^4}\, ,\label{eq:modelsc}
\end{eqnarray}
\end{subequations}
to describe our numerical observations in Fig.~\ref{fig:Asymp} based
on the Monte Carlo simulations. The model that best describes our
observations allows us to determine, with a certain degree of
confidence, the performance of non-Gaussian adaptive strategies, and
compare them with the best Gaussian strategy, Eq.~\eqref{eq:MKII_var},
and the canonical phase measurement, Eq.~\eqref{eq:CPM_var}.

We discriminate among plausible candidate models using the technique of backward
elimination \cite{Young2017}. We start with the candidate $\widehat{y}_1$,
Eq.~\eqref{eq:modelsa}, and test the deletion of each variable $A_i$ using the
$p$-value of a hypothesis testing procedure ($H_0: \, A_i = 0$, $H_A: \, A_i
\neq 0$). Given that $p> 0.1$ for $A_2$ and $p< 0.001$ for $A_1$ and
$A_3$, we conclude, with confidence larger than $99\%$ , that the model
reflecting the behavior of the data presented in Fig.~\ref{fig:Asymp} is
$\widehat{y}_3$, Eq.~\eqref{eq:modelsc}.

\begin{figure*}[hbt!]
  \centering
  \includegraphics[width=\textwidth]{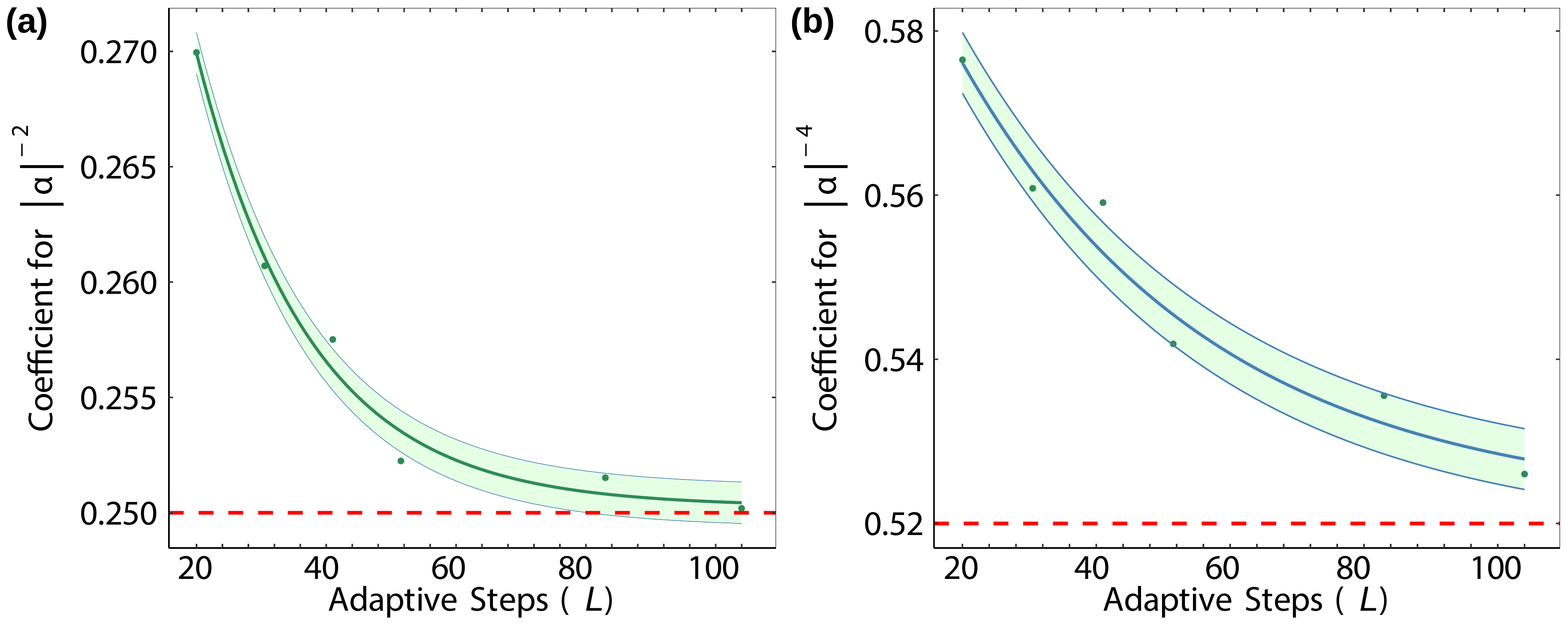}
  \caption{ \textbf{Coefficients of model $\widehat{y}_3$ in
      Eq.~\eqref{eq:modelsc} for the Holevo Variance $\text{Var}_{\text{H}}[
      \widehat{\phi} ]$, as a function of $L$}. Note that in the limit of $L \to
    \infty$ the coefficient $A_1$ for the term $1/\lvert \alpha \rvert^2$ tends
    to $0.25$, and $A_3$ of term $1/\lvert \alpha \rvert^4$ tends to $0.52$.
    Curves are fits to an exponential model $A_i=D_i\exp(-E_i L)+F_i$ with
    coefficients $(D,E,F,\mathrm{RSE})=(0.065 \pm 0.013,0.060\pm 0.010, 0.250\pm
    0.001, 1.87\times 10^{-7})$ for the left panel, and
    $(D,E,F,\mathrm{RSE})=(0.095\pm 0.015, 0.279\pm 0.119, 0.522\pm
    0.010,1.82\times10^{-5})$ for the right panel. The shaded regions represent
    a 1-$\sigma$ standard deviation.}
  \label{fig:Asymp2}
\end{figure*}

To find $A_1$ and $A_3$ in the limit $L\to\infty$ in model $\widehat{y}_3$, we fit
the Holevo variance as a function of $|\alpha|^2$ for $\lvert {\alpha} \rvert^2
>5$ to the model $\widehat{y}_3$ for different values of $L$. Given a number of
adaptive steps $L$, each fitting provides a set of coefficients $\left\{ A_1,
  A_3 \right\}$. Hence, to obtain the trend of each coefficient $A_1$ and $A_2$
as $L$ increases, we fit them to an exponential model of the form $A_i=D_i\exp
\left( -E_i L \right)+F_i$. Fig.~\ref{fig:Asymp2} shows the coefficients $A_1$
and $A_3$ as a function of $L$. Adjusting this exponential model and observing
that in the limit $L \gg 1$ $A_i\rightarrow F_i$, we obtain $A_1=0.250 \pm
0.001$ and $A_3= 0.52\pm 0.01$. Therefore, we conclude with a $99\%$ confidence
level that the Holevo variance for the adaptive non-Gaussian strategy in the
limit $L\to\infty$ for large $|\alpha|$ is:

\begin{equation}
  \label{eq:modelstrb}
  \text{Var}_{\text{H}}\left[ \widehat{\phi} \right] \approx \frac{0.250\pm 0.001}{\lvert
    \alpha  \rvert^2 } + \frac{0.520\pm 0.010}{\lvert \alpha  \rvert^4}\, .
\end{equation}

This equation is the main result of this work, also shown in Eq.
(\ref{eq:model}). We observe that the Holevo variance for the adaptive
non-Gaussian strategy has a similar dependance with $|\alpha|$ as the
canonical phase measurement, differing only in the scaling of the
correction term of order $1/|\alpha|^4$, see Eq. \eqref{eq:CPM_var}.
Moreover, we note that the best Gaussian strategy known to date, the
MKII adaptive homodyne, has a correction term of order $1/|\alpha|^3$
\cite{Wiseman1995}, see Eq. \eqref{eq:MKII_var}. Then, for large
$|\alpha|$, the non-Gaussian estimation strategies show a much better
scaling in the Holevo variance than the best known Gaussian strategy,
and closely follows the canonical phase measurement. Furthermore,
these non-Gaussian phase estimation strategies can be implemented with
current technologies \cite{DiMario2020}, and our work demonstrates
their superior performance over all the physically realizable
strategies for single-shot phase estimation of coherent states
reported to date.

The optimized non-Gaussian adaptive strategies based on photon counting analyzed
in this work are not the only possible strategies for single-shot phase
estimation, and there could be other strategies based on photon counting with
better performances. In this work, we studied estimation strategies using cost
functions that are consistent with D-optimal designs. However, there may be
other cost functions that could provide a further improvements to non-Gaussian
strategies. Moreover, we note that the relation in Eq.~(\ref{eq:beta_opt_hat})
between the phase and the magnitude of the displacement field was used to obtain
identifiable likelihood functions and ensure efficient estimators in the
asymptotic limit. While we chose the relation in Eq.~(\ref{eq:beta_opt_hat})
because it maximizes the Fisher information, there is no mathematical proof that
this choice is optimal, or that other choices for this relation would not
provide higher sensitivities. Finally, in the presented adaptive non-Gaussian
strategies for phase estimation, the displacement operations were optimized in
every adaptive step at a time, i. e. using local optimizations in the adaptive
steps. We note, however, that local optimal strategies do not necessarily ensure
global optimality \cite{ferdinand2017multi}. Strategies with global
optimizations, where all adaptive steps are optimized simultaneously, could
probably lead to better performances. However, the computational overhead
required for performing global optimizations beyond $L=10$ adaptive steps
prevents us from being able to investigate global optimized strategies. To
overcome these limitations, future investigations could make use of machine
learning methods, such as neural networks and reinforcement learning
\cite{dimario21}, to lower the complexity of these calculations.
\\

\section{\label{sec:Discussion} Discussion }

Non-Gaussian measurement strategies for phase estimation approaching the quantum
limits in sensitivity, as set by the canonical phase measurement, can become a
tool for enhancing the performance of diverse protocols in quantum sensing,
communications, and information processing. Optical phase estimation approaching
the quantum limit can be used to prepare highly squeezed atomic spin states
based on measurement backaction \cite{DEUTSCH2010681} for quantum sensing and
metrology \cite{RevModPhys.90.035005}; enhance phase contrast imaging of
biological samples with optical probes at the few photon level to avoid
photodamage and ensure integrity of the sample; and to enhance fidelities in
quantum communications with phase coherent states that require phase tracking
close to the quantum level between receiver and transmitter with few-photon
pulses \cite{PhysRevX.5.041009, PhysRevX.5.041010}, while allowing for quantum
receivers to decode information encoded in coherent states below the quantum
noise limit \cite{dimario18b, ferdinand2017multi,becerra2015photon}.

As a direct application for quantum information processing, non-Gaussian
measurements based on photon counting and displacement operators can be used for
full reconstruction of quantum states with on-off detectors
\cite{PhysRevA.80.022114} and PNR detectors \cite{Nehra:19} in a multi-copy
state setting by sampling phase space with non-Gaussian projections. The
theoretical methods to assess the performance of adaptive non-Gaussian measures
for phase estimation described in this work could be used to study methods for
adaptive quantum tomography \cite{PhysRevA.85.052120, Granade_2017} based on
photon counting for high dimensional quantum states, and investigate their
asymptotic advantages over adaptive homodyne tomography \cite{PhysRevA.60.518}.


From the practical point of view, our work provides insight into the design of
practical, highly efficient measurement strategies for phase estimation based on
photon counting. It shows that non-Gaussian strategies optimizing any cost
function within the family of D-optimal designs are equally efficient, and
demonstrates the advantages of higher photon number resolution in the strategies
to reduce estimation errors. This understanding allows the experimenter to chose
cost functions that can be efficiently calculated and optimized to achieve
higher convergence rates, while selecting a PNR given the desired/target error
budget in the estimation problem for specific applications. This knowledge will
be critical for the design and development of future sensors based on photon
counting for diverse applications in communication, phase-contrast imaging,
metrology, and information processing.


In conclusion, we investigate a family of non-Gaussian strategies for
single-shot phase estimation of optical coherent states. These strategies are
based on adaptive photon counting measurements with a finite number of adaptive
steps implementing coherent displacement operations, optimized to maximize
information gain as the measurement progresses \cite{DiMario2020}. We develop a
comprehensive statistical analysis based on Bayesian optimal design of
experiments that provides a natural description of adaptive non-Gaussian
strategies. This theoretical framework gives a fundamental understanding on how
to optimize these strategies to enable efficient estimators with a high degree
of convergence towards the ultimate limit, the Cramér Rao lower bound.

We use numerical simulations to show that optimized adaptive non-Gaussian
strategies producing an asymptotically efficient normal estimator achieve a much
higher sensitivity than the best Gaussian strategy known to date, which is based
on adaptive homodyne \cite{Wiseman1995}. Moreover, we show that the Holevo
variance of the estimator for the adaptive non-Gaussian strategy has a similar
dependence as the canonical phase measurement in the asymptotic limit of large
photon numbers, differing by a scaling factor in the second-order correction
term. Our work complements the work in single-shot phase measurements for
single-photon wave packets in two dimensions \cite{Martin2019} using quantum
feedback with Gaussian measurements, and paves the way for the realization of
optimized feedback measurements approaching the canonical phase measurement for
higher dimensional states based on non-Gaussian operations.
\\
\\
\textbf{DATA AVAILABILITY}
\\
The data that support the findings of this study are available from the authors
upon request.
\\
\\

\textbf{CODE AVAILABILITY}
\\
Our numerical results have been implemented using MATLAB and Julia language. A
functions to reproduce the numerical results can be publicly found in the
repository \cite{Repo}. To obtain a set of estimations produced by the non
Gaussian strategy, use the program: \verb+ Bootstrap_phase_estimation.jl+,
setting the desired values of $L$, Bootstrap size, MPN, in the
\verb+Adapt_Steps+, \verb+Boot_Reps+, and \verb+MPNS+ variables.
\\
\\

\textbf{\label{sec:Ack} ACKNOWLEDGMENTS}
\\
We thank Laboratorio Universitario de Cómputo de Alto Rendimiento (LUCAR) of
IIMAS-UNAM for their service on information processing. This research was
supported by the Grant No. UNAM-DGAPA-PAPIIT IG101421 and the National Science
Foundation (NSF) grans \# PHY-1653670 and PHY-2210447.
\\
\\
\textbf{\label{sec:Cont} AUTHOR CONTRIBUTIONS}
\\
F.E.B. and P.B. conceived the idea and supervised the work. M.A.R. and P.B.
conducted the theoretical and statistical study. M.T.D. and M.A.R. contributed
in the code for simulations. All authors contributed to the analysis of the
theoretical and numerical results and contributed to writing the manuscript.
\\
\\

\noindent
\textbf{COMPETING INTERESTS}
\\
The authors declare that there are no competing interests.

\noindent



\bibliographystyle{naturemag}
\bibliography{References}

\end{document}